# THE RESULT FOR THE GRUNDY NUMBER ON P4-CLASSES

Ali Mansouri[1] and Mohamed Salim bouhlel [2]

[1]Department of Electronic Technologies of Information and Telecommunications Sfax, Tunisia
[2]Department of Electronic Technologies of Information and Telecommunications Sfax, Tunisia

*ABSTRACT*

*Our work becomes integrated into the general problem of the stability of the network ad hoc. Some, works attacked (affected) this problem. Among these works, we find the modelling of the network ad hoc in the form of a graph. We can resume the problem of coherence of the network ad hoc of a problem of allocation of frequency*

*We study a new class of graphs, the fat-extended P4 graphs, and we give a polynomial time algorithm to calculate the Grundy number of the graphs in this class. This result implies that the Grundy number can be found in polynomial time for many graphs*

*KEYWORDS*

*Graph Theory, Grundy number, P4-classes, Total graph, Graph coloring, Wireless Network, Mobile Network,*

## 1. INTRODUCTION

We denoted a graph G by $G = (V, E)$ and an order $= v_1, \ldots, v_n$ over V, for coloring the vertices of G, we used the polynomial algorithm, we affected to $v_i$ the minimum positive integer that was not affected to its neighborhood in the set $\{v_1, \ldots, v_{i-1}\}$[10].

The execution of this algorithm for a graph give a coloring that we can gives name as a greedy coloring.

The greedy coloring of a graph G is characterized by the maximum number of colors of, over all the orders of V (G) is the Grundy number of G; we represented by (G) the grundy number of graph. [10]. the Grundy number is considered NP-complete [9]. We can show a graph G and an integer r it is a coNP-complete problem to decide if the grundy number of G respect the following condition (G) + r [9] or if (G) r × (G) or if (G) c × (G) [2]. In the next section, we study the polynomial time algorithms to determinate the Grundy number of many classes of graphs like the cographs [4], the trees [5] and the k-partial trees [8].

## 2. RELATED WORKS

This parameter was introduced by Christen and Selkow [11] in 1979. They proved that determining the Grundy number is NP-complete for general graphs (also studied by McRae in 1994 [AMC94]). In [SST84], Hedetniemi et al. gave a linear algorithm for the Grundy number of a tree and established a relation between the chromatic number, the Grundy number and the achromatic number: (G)   (G)   (G), where the achromatic number   (G) is the maximum number of colors used for a proper coloring of G such that each pair of colors appears on at least one edge of G.

In 1997, Telle and Proskurowski [12] gave an algorithm for the Grundy number of partial k-trees in O ($n3k2$) and bounded this parameter for these graphs by the value $1 + k \log_2 n$, where n is the graph order. In 2000, Dunbar et al. used the Grundy number to bound new parameters that they introduced in [13], the chromatic and the achromatic numbers of a fall coloring.

Recently, Germain and Kheddouci studied in [14], the Grundy coloring of power graphs. They gave bounds for the Grundy number of the power graphs of a path, a cycle, a caterpillar and a complete binary tree. Such colorings are also explored for other graphs like chessboard graphs [15].

Hedetniemi et al. (1982 determinate an algorithm witch can gives a solution for  this problem on trees; and an algorithm on partial k-trees gives by Telle and Proskurowski (1997). The problem was shown to be NP-complete on directed graphs Christen and Selkow (1979).

## 3. FAT EXTENDED P4-LADEN GRAPHS

In this section, we give some definitions. In first time, we denote G by G = (V, E) is a graph and S is a subset of V (G) [10].The subgraph of G is represented by G[S]. The module of a graph G is denoted or represented by M [10]. The sets V and {x}, for every x   V, are trivial modules, the last one being called as a singleton module.

We called graph prime if this graph have all of modules are trivial. M is considered like a strong module of G if this condition is approved "|for every module M  of G, either M   M =    or M   M or M    M" [10].

We denoted by  r an internal node of T(G), we denoted by M(r) the set of leaves of the sub tree of T(G) , also we denoted by V (r) = {$r_1$, . . . , $r_k$} the children of r in T(G). After that, If G [M(r)] is not connected, then we called r a parallel node and the graph G [M ($r_1$)],. . . ,G M ($r_k$)] are define like its components.  Now , If the graph $G^-$ [M(r)] is not connected , so we called r a set of  set node , $G^-$ [M ($r_1$)],. . . ,$G^-$ [M($r_k$)] define the  components of graph. Finally, r is define like a neighbourhood node and M ($r_1$), . . . , M($r_k$) is called the unique set of maximal submodules of M(r) , if the graphs G [M(r)] is connected at $G^-$ [M(r)]. we denoted, by quotient graph of r,  the G(r), is G [$v_1$, . . . , $v_k$], where vi    M($r_i$), for 1    i    k. r is called a fat node, with simple condition if  M(r) is not a singleton module.

We called spider graph, a graph have vertex set can be devised into three sets S, K and R
 We denoted a stable set by S, K represents a clique, the set of vertices of R are adjacent to all the vertices of K and is not adjacent of the vertices of S

A graph is called a graph split if it is {C5, C4, C¯4}-free. [10] A pseudo-split graph is defined as a {C4, C¯4}-free graph. Also, a split graph G = (S ∪ K, E), its vertex set can be devised into three disjoint sets S (G), K (G) and R (G)

There is condition must be respected "S(G) ⊂ S and every vertex s ∈ S(G) is not adjacent to at least one vertex in K, K(G) is the neighborhood of the vertices in S(G) and R(G) = V (G)\S(G) ∪ K(G). [10]

Giakoumakis [3] defined a graph G as extended P4-laden graphs if, for all H ⊂ G such that |V (H)| ≤ 6, then the following statement is true: if H contains more than two induced P4's, then H is a pseudo-split graph. We can define an extended P4-laden graph by its modular partition tree

**Theorem 1** [3] we denote a graph G by G = (V, E), we denote by T (G) the modular partition tree and by r the any neighborhood node of T (G),

Then G is extended P4-laden if and only if G(r) is isomorphic to:
(i) a P5 or a P¯5 or a C5, and each M(ri) is a singleton module; or
(ii) a spider H = (S ∪ K ∪ R, E) and each M(r$_i$) is a singleton module, except the one corresponding to R and eventually another one which may have exactly two vertices; or
(iii) a split graph H, whose modules corresponding to the vertices of S(H) are independent sets and the ones corresponding to the vertices of K(H) are cliques.

We say that a graph is fat-extended P4-laden if its modular decomposition satisfies the Theorem 1, except in the first case, where G(r) is isomorphic to a P5 or a P¯5 or a C5, but the maximal strong modules M(r$_i$), 1 ≤ i ≤ 5, of M(r) are not necessarily singleton modules.

## 4. GRUNDY NUMBER ON FAT EXTENDED P4-LADEN GRAPHS

From now, let G = (V, E) be a fat-extended P4-laden graph and T (G) its modular decomposition tree. The modular partition tree T (G) can be determinate in polynomial time [7], we give an algorithm to determinate Γ(G). We know that the Grundy number of the leaves of T (G) is equal to one and we show in this section how to determine the Grundy number of G[M(v)], for every inner node v of T(G), based on the Grundy number of its children.

First, observe that for every series node (resp. parallel node) v of T (G), the Grundy number of G[M(v)] is equal to the sum of the Grundy number of its children (resp. the maximum Grundy number of its children) [4]. Thus, we only need to prove that the Grundy number of G [M (v)] can be found in polynomial time when v is a neighborhood node of T (G).

The following result is a simple generalization of a result due to Asté etal. [2] for the Grundy number of lexicographic product of graphs:

**Proposition 2:** Let G, H$_1$, . . . , H$_n$ be disjoint graphs such that n = |V (G)| and let V (G) = {v$_1$, . . . , v$_n$}. If G is the graph obtained by replacing v$_i$ ∈ V (G) by H$_i$, then for every greedy coloring of G at most Γ(H$_i$) colors contain vertices of the induced subgraph G [H$_i$] ⊂ G , for all i ∈ {1, . . . , n}.

**Lemma 3:** Let v be a neighborhood node of T(G) isomorphic to a P5 or a C5 or a C¯5, v$_1$, . . . , v$_5$ be the children of v and Γ$_i$ be the Grundy number of G[M(v$_i$)], 1 ≤ i ≤ 5. Then Γ(G [M (v)]) can be found in constant time.

**Proof:** (Sketch) Without loss of generality, suppose that $v_1, \ldots, v_5$ label the children of v as depicted in Figure 1 and Hi = (G[M($v_i$)]). In order to simplify the notation, denote by $\pi_i$ an ordering over M ($v_i$) that induces a greedy coloring with $\Gamma_i$ colors, $1 \le i \le 5$.

We calculate $\Gamma$(G [M (v)]) by verifying all the possible configurations for a greedy $\Gamma$(G [M (v)])-coloring and by returning the greater value found between all the cases. Suppose that G (v) is isomorphic to a P5. Let S = {$S_1, \ldots, S_k$} be a greedy $\Gamma$(G[M(v)])-coloring of G[M(v)].

We claim that if there exists a vertex u $\in$ V ($H_1$) colored by $S_k$, then $\Gamma$(G [M (v)]) = $\Gamma_1 + \Gamma_2$. This fact holds because combining the observation that u has at least one vertex colored by $S_i$, for all i $\in$ {1, ..., k −1}, with the proposition 2, we conclude that $\Gamma$(G[M(v)]) $\le \Gamma_1 + \Gamma_2$. On the other hand, if we consider any ordering $\pi$ over G [M (v)] that has starts with $\pi_1$ and $\pi_2$, we see that the first-fit algorithm over this order will produce a greedy coloring with at least $\Gamma_1 + \Gamma_2$ colors. Using the symmetry, we can also prove that if u $\in$ V ($H_5$), then $\Gamma$(G [M (v)]) = $\Gamma_4 + \Gamma_5$.

All the other cases use similar arguments, that is, by finding an upper bound based on the position of a vertex colored $S_k$ and a lower bound based in an ordering over M (v). The cases where G (v) is isomorphic to C5 or $\overline{P}$5 are also proved by using similar arguments [10].

**Lemma 4 :** Let v be a neighborhood node of T (G) isomorphic to a spider H = (S $\cup$ K $\cup$ R, E), fr be its child corresponding to R, f2 be its child corresponding to the module which has eventually two vertices and $\Gamma$(R) be the Grundy number of G[M(fr)]. Then $\Gamma$(G [M (v)]) can be found in linear time.

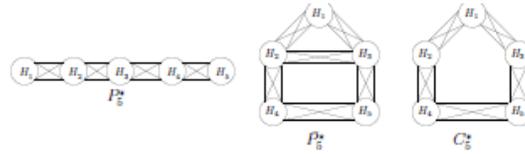

Figure 1: Fat neighborhood nodes.

**Proof:** If M (f2) is singleton module, then G [M (v)] is a spider. In this case, we cannot have two colors $S_i$ and $S_j$, j > i, such that both contain only vertices of S. For otherwise, since S is a stable set, the vertices colored $S_j$ would not any neighbor colored $S_i$, a contradiction. Thus, $\Gamma$(G [M (v)]) $\le$ 1 + |K| + $\Gamma$(R). If R = $\emptyset$, then an ordering over M (v) such that all the vertices of S come before the vertices of K induces a greedy coloring with $\Gamma$(G[M(v)]) = 1 + |K| colors. If R $\ne \emptyset$, we will prove that $\Gamma$(G [M (v)]) $\ge$ |K| + $\Gamma$(R). Observe first that there is at least one color Si occurring in R. Consequently, Si does not occur in K. Thus, there is no order over M(v) whose greedy coloring returns a color $S_j$ containing only vertices of S, because a vertex of S colored $S_j$ would not be adjacent to a vertex colored $S_i$. On the other hand, if $\pi_R$ is an ordering that induces a greedy $\Gamma$(R)-coloring of R, then any ordering over M(v) starting by $\pi_R$ induces a greedy coloring with at least |K| + $\Gamma$(R) colors.

The case where M (f2) is not a singleton module is proved using similar arguments. [10]

If G (v) is a split graph H and the factors corresponding to vertices of S (H) are independent sets and the ones corresponding to vertices of K (H) are cliques, then we can use the same arguments of Lemma 3.3 observing that S, K and R correspond to S (H), K (H) and R (H), respectively .

**Theorem 5:** If $G = (V, E)$ is a fat-extended P4-laden graph and $|V| = n$, then (G) can be found in $O(n_3)$.

**Proof:** The algorithm calculates (G) by traversing the modular decomposition tree of G in a postorder way and determining the Grundy of each inner node of T (G) based on the Grundy number of the leaves. The modular decomposition tree can be found in linear time, the postorder traversal can be done in O (n2) and the Grundy number of each inner node can be found in linear time, because of Lemmas 3and 4 and the results of Gyarf´as and J.Lehel [4] for cographs.

**Corollary 6:** Let G be a graph that belongs to one of the following classes:

P4-reducible, extended P4-reducible, P4-sparse, extended P4-sparse, P4-extendible, P4-lite, P4-tidy, P4-laden and extended P4-laden. Then, (G) can be found in polynomial time.

**Proof:** According to definition of these classes [6], they are all strictly contained in the fat-extended P4-laden graphs and so the corollary follows.

## 5. CONCLUSIONS

In the conclusion, we determinate, in first time, a new class of graphs, this class is called the fat-extended P4-laden graphs, in second time; we calculate the Grundy number of the graphs in this class by using a polynomial time algorithm. Finally, we can say that the Grundy number can be calculated of each node in linear time, so, the total Grundy number of graph can be determinate in polynomial time.

## ACKNOWLEDGEMENTS

I think Mr. Julio Cesar Silva Araujo for his collaboration.

## REFERENCES

[1] J. C. Araujo., "Coloring graphs",(2009). Master in Science Dissertation, in: ww.lia.ufc.br/˜juliocesar/dissertacao.pdf (in Portuguese)".
[2] M., F. Havet & C. Linhares Sales, (2008) "Grundy number and lexicographic product of graphs" in: International Conference on Relations, Orders and Graphs and their Interaction with Computer Science
[3] V. Giakoumakis., (1996) "p4-laden graphs: A new class of brittle graphs", Information Processing Letters 60 pp, 29–36.
[4] A, Gyarf´as. & J. Lehel, (1988) "On-line and first fit colorings of graphs", Journal of Graph Theory pp. 217–227.
[5] T. Beyer S. M. & S. T. Hedetniemi, (1982) "A linear algorithm for the grundy number of a tree", Congressus Numerantium 36, pp. 351–363.
[6] Pedrotti, V, (2007) "Decomposi¸c˜ao modular de grafos n˜ao-orientados," Master's thesis, Universidade Estadual de Campinas
[7] Tedder, M., D. Corneil, & C. Paul, (2007)." Simple, linear-time modular decomposition".
[8] J, Telle A. & A. Proskurowski, (1997) "Algorithms for vertex partitioning problems on partial k-trees", SIAM Journal on Discrete Mathematics 10 pp. 529–550.
[9] M, Zaker, (2005) "Grundy number of the complement of bipartite graphs" Journal of Combinatorics 31 pp. 325—330.
[10] J. C. Silva Ara´ujo, C. L. Sales, (2009) "Grundy number on P4-classes", Electronic Notes in Discrete Mathematics, Volume 35, 1, Pages 21–27.
[11] M. Selkow C. & A. Christen (1979) " Some perfect coloring properties of graphs", Journal of Combinatorial Theory B27 ,49-59.

[12] A. Proskurowski, & J.A. Telle, (1997), "Algorithms for vertex partitioning problems on partial k-trees", SIAM Journal on Discrete Mathematics 10(4), 529-550.
[13] S.M. Hedetniemi, J.E. Dunbar, D.P. Jacobs & J. Knisely, (2000), "Colorings of Graphs", Journal of Combinatorial Mathematics and Combinatorial Computing 33 ,257-273.
[14] C. Germain and H. Kheddouci, "Grundy coloring for power caterpillars", (2000), Proceedings of International Optimization Conference INOC 2003, 243-247, Evry/Paris France.
[15] J. Rhyne&C. Parks (2002) "Grundy Coloring for Chessboard Graphs", Seventh North Carolina Mini-Conference on Graph Theory,